\documentclass[twocolumn,superscriptaddress,aps,prl]{revtex4}
\usepackage{graphicx}
\usepackage[novolumeabbr]{unitsdef} 
\newunit{\invcm}{\centi\meter\unitsuperscript{-1}}

\begin{document}
\bibliographystyle{prsty}

\title{Terahertz conductivity of twisted bilayer graphene}
\author{Xingquan Zou}
\author{Jingzhi Shang}
\author{Jianing Leaw}
\author{Zhiqiang Luo}
\author{Liyan Luo}
\author{Chan La-o-vorakiat}
\author{Liang Cheng}
\author{S. A. Cheong}
\affiliation{Division of Physics and Applied Physics, School of Physical and Mathematical Sciences, Nanyang Technological University, 637371, Singapore}
\author{Haibin Su}
\affiliation{Division of Materials Science, School of Materials Science and Engineering, Nanyang Technological University, 639798, Singapore}
\author{Jian-Xin Zhu}
\affiliation{Theoretical Division and Center for Integrated Nanotechnologies, Los Alamos National Laboratory, Los Alamos, New Mexico 87545, USA}
\author{Yanpeng Liu}
\author{Kian Ping Loh}
\affiliation{Department of Chemistry, National University of Singapore, 3 Science Drive 3, 117543, Singapore}
\author{A. H. Castro Neto}
\affiliation{Graphene Research Centre and Physics Department, National University of Singapore, 6 Science Drive 2, 117546, Singapore}
\author{Ting Yu}
\author{Elbert E. M. Chia}
\affiliation{Division of Physics and Applied Physics, School of Physical and Mathematical Sciences, Nanyang Technological University, 637371, Singapore}

\begin{abstract}
Using terahertz time-domain spectroscopy, the real part of optical conductivity [$\sigma_{1}(\omega)$] of twisted bilayer graphene was obtained at different temperatures (10 -- 300~K) in the frequency range 0.3 -- 3~THz. On top of a Drude-like response, we see a strong peak in $\sigma_{1} (\omega)$ at $\sim$2.7~THz. We analyze the overall Drude-like response using a disorder-dependent (unitary scattering) model, then attribute the peak at 2.7~THz to an enhanced density of states at that energy, that is caused by the presence of a van Hove singularity arising from a commensurate twisting of the two graphene layers.
\end{abstract}

\maketitle
Compared to single-layer graphene (SLG), where there are two non-equivalent lattice sites ($A$ and $B$), bilayer graphene (BLG) has two SLGs stacked in the third direction. In the most common Bernal ($AB$) stacking of BLG, adjacent layers are rotated by 60$^{\circ}$, so that the $B$ atoms of layer 2 ($B^{\prime}$) sits directly on top of $A$ atoms in layer 1 ($A$), and $B$ and $A^{\prime}$ atoms are in the center of the hexagons of the opposing layer. Electrons can then hop between these two $A$ sites with a hopping energy $t_{\perp}$. In the undoped case, though both SLG and BLG are gapless semi-metals, carriers in SLG exhibit linear dispersion, while those in BLG show quadratic dispersion. An energy gap in SLG opens up due to finite geometry effects, but its control has proven to be unreliable \cite{Nilsson08a}. On the other hand, the electronic gap in BLG can be reliably opened and controlled by an applied electric field, shown theoretically and demonstrated experimentally \cite{McCann06a,McCann06b,Castro07a,Oostinga07}, and promises interesting applications. Both SLG and BLG however, are sensitive to disorder. Hence, to realize graphene-based optoelectronic devices, an understanding of the temperature and disorder effects in the transport and spectroscopic properties of BLG is needed. Temperature and disorder-dependent conductivity of BLG have been derived theoretically \cite{Dahal08,Nilsson08a}. Experimentally, spectroscopies (from terahertz (THz) to visible) and ultrafast dynamics of various flavors of graphene have been reported, such as SLG, few and many-layer graphene, and graphite \cite{Choi09a,Peres06a,Zou10a,Shang10a,Kuzmenko08a}. For example, Fourier-transform infrared spectroscopy (FTIR) on large-area SLG grown by chemical vapor deposition (CVD) revealed a Drude-like frequency dependence of the spectral density from THz to mid-infrared at different carrier concentrations \cite{Horng11a}. In addition, graphene plasmons, which lie in the THz range, are strongly coupled to the interband electronic transitions and decay by exciting interband electron-hole pairs \cite{Rana08a}. Hence knowledge of graphene's electromagnetic response, as a function of disorder, in the THz frequency range is critical for applications such as graphene-based THz oscillators \cite{Dubinov09a}.

\begin{figure} \centering \includegraphics[width=7cm,clip,angle=270]{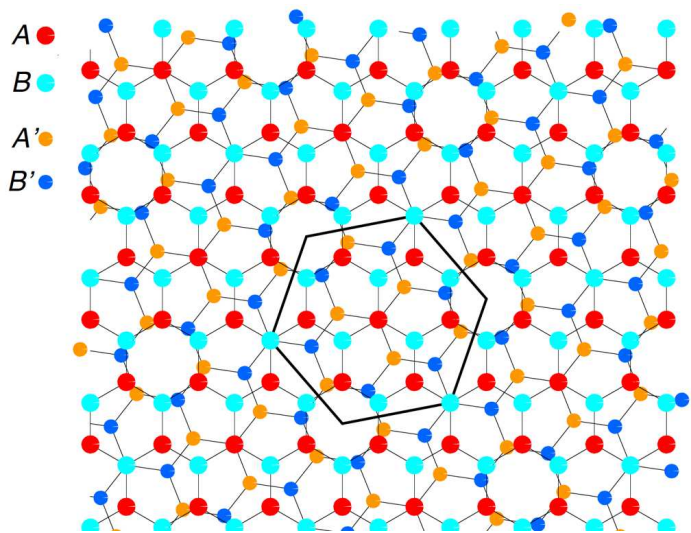}
\caption{(Color online) Atomic arrangement of atoms in BLG, for twisting angle $\theta$=21.8$^{\circ}$ (corresponding to $i=1$). $A$ ($A^{\prime}$) and $B$ ($B^{\prime}$) are the sublattices of the first (second) layer. The black hexagon depicts the unit cell of the twisted BLG. Compare with Bernal (AB)-type stacking, where $\theta$=60$^{\circ}$ (i.e. $i=0$).}
\label{fig:Twisted}
\end{figure}

BLG grown by CVD also has a great tendency to twist. A typical 10~mm x 10~mm piece of CVD-BLG has been shown to be a collection of crystallites of twisted BLG with a distribution of different twisting angles \cite{Li10a,Reina09a}. Twisting occurs when there is rotation between the top and bottom layers of BLG (see Fig.~\ref{fig:Twisted}). When there is rotation through a (twisting) angle $\theta$ about an $A$ ($B^{\prime}$) site in BLG, only a discrete set of commensurate angles is allowed~\cite{Santos07a}:
\begin{equation}
\cos (\theta_i) = \frac{3i^2 + 3i + 1/2}{3i^2 + 3i + 1},
\label{eqn:theta}
\end{equation}
where $i = 0, 1, 2, ...$. Such rotation between graphene layers have been observed as a Moir$\acute{e}$ pattern on graphite surfaces \cite{Rong93a}, and recently in BLG \cite{Li10a}. Such twisting causes van Hove singularities (VHS) to develop near the Fermi energy, with the VHS energy scale being a strong function of $\theta$, resulting in an enhanced density of states at those energies \cite{Santos07a}. Such enhancement in the density of states should show up in the conductivity spectrum. For example, for large twisting angles of 7.5$^{\circ}$, 13.7$^{\circ}$ and 54.6$^{\circ}$, anomalies in the real conductivity $\sigma_{1}(\omega)$ were seen in the visible region using contrast spectroscopy \cite{Wang10a}. However, they have not been demonstrated in the THz regime. An accurate characterization of electrical and optical conductivities at THz frequencies of BLG, as a function of temperature, disorder \textit{and twisting}, is thus needed, but has not been reported.

In this Letter, we present THz time-domain spectroscopy (THz-TDS) studies of twisted BLG at different temperatures (10 K -- 300 K), to study its frequency-dependent far-infrared conductivity. On top of a Drude-like response, we see a peak in the real conductivity. The overall Drude shape was analyzed using a disorder-dependent model, while the conductivity peak at 2.7~THz was attributed to an enhanced density of states at that energy, that is caused by the presence of low-energy VHS arising from a commensurate twisting of the graphene layers relative to each other.

\begin{figure} \centering
\includegraphics[width=8cm,clip]{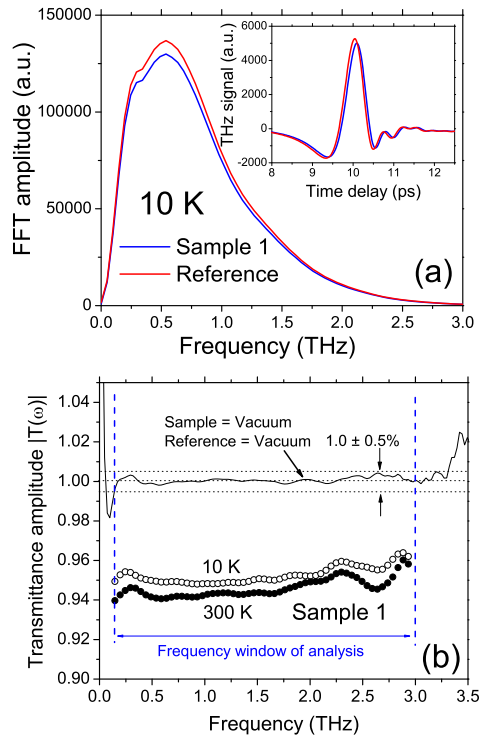}
\caption{(Color online) (a) Amplitude spectra of Sample 1, obtained from the Fourier Transform of the main pulse in the inset. Inset: THz-TDS signal of Sample 1 and reference at 10~K. (b) Amplitude of complex transmission coefficient (or transmittance), at 10~K and 300~K. The solid line indicates the transmission amplitude when both sample and reference positions are vacuum.}
\label{fig:TDS}
\end{figure}

The samples studied here are large-scale BLG grown by CVD and  deposited on z-cut quartz. Both contrast and absorption spectroscopies confirmed the sample to be a BLG film \cite{Ni07}. Our experimental set-up performs an average over the entire area of the sample. Nevertheless, our data were able to discern the feature arising from $\theta_{i=28} = 1.161^{\circ}$, on top of a broad background produced by the disorder in the sample, as will be discussed later in this Letter.

The transmission THz spectra of the BLG were measured using a conventional THz-TDS system (TeraView Spectra 3000) with a Janis ST-100-FTIR cryostat. THz-TDS has proven to be a very useful noncontact technique to study material properties such as dielectric response, complex conductivity and refractive index in the far-infrared range without the need for Kramers-Kronig analysis \cite{Grischkowsky90a,Baxter06a}. The THz wave was generated and detected by photoconductive antennas fabricated on low-temperature-grown GaAs films. The aperture diameter is 7~mm, enabling accurate measurements of the low-frequency spectral components of the THz wave. For each sample or reference run, 900 THz traces were taken over 180 seconds. The sample holder was moved from the sample to the reference position (and vice versa) by means of a vertical motorized stage with 2.5~$\mu$m resolution. The time-domain electric fields of a THz pulse transmitted through Sample 1 (${\tilde E_s}(t)$ --- BLG deposited on 1-mm thick z-cut quartz substrate from CrysTec, Germany), as well as through the reference (${\tilde E_r}(t)$, bare z-cut quartz substrate) are shown in the inset of Fig.~\ref{fig:TDS}(a). Before BLG deposition, the substrates for sample and reference were carefully characterized by THz-TDS --- their phase difference yields the thickness difference $\Delta L$ between the two substrates, which must be taken into account in our subsequent analysis \cite{Kadlec09}. After the main pulse, a weaker pulse (etalon) appears due to multiple reflections in the z-cut quartz substrate. Since the main pulse and etalon pulse are well separated in the time domain, we truncate the time-domain data to remove the etalon pulse. Subsequent data analysis was performed only on the main pulse without loss of validity. Fast Fourier Transform (FFT) was then performed to obtain the amplitude and phase at different spectral components of the THz pulse. The FFT amplitude spectrum of the main pulse is shown in Fig.~\ref{fig:TDS}(a). The absorption of the THz pulse by the BLG is obvious, even though the sample is of atomic-scale thickness. Figure~\ref{fig:TDS}(b) shows the amplitude of the experimental transmission coefficient (or transmittance) $T(\omega)$, defined as the ratio between complex spectral field of the sample $\tilde E_{s}(\omega)$ and reference $\tilde E_{r}(\omega)$, for the BLG sample at 10~K and 300~K. For both temperatures $|T(\omega)|$ is almost frequency-independent with the value $\sim$95$\percent$. In the same figure is $|T(\omega)|$ when both sample and reference are vacuum --- in this case $|T(\omega)|$ deviates only 0.5$\percent$ away from unity in the frequency range 0.3 -- 3.0~THz, which will be the frequency window of our analysis.

Theoretically, for a sample grown on a substrate, $T(\omega)$ can be written as \cite{Duvillaret96a}
\begin{equation}
 T(\omega) =\\
  \frac{{2\tilde n({{\tilde n}_{sub}} + 1)\exp [i\omega d(\tilde n - 1)/c]\exp [-i\omega \Delta L({{\tilde n}_{sub}} - 1)/c]}}{{(1 + \tilde n)(\tilde n + {{\tilde n}_{sub}}) + (\tilde n - 1)({{\tilde n}_{sub}} - \tilde n)\exp [2i\omega d\tilde n/c]}} \label{eqn:Tw}
 \end{equation} where $\tilde n$ and ${\tilde n_{sub}}$ are the complex refractive indices of BLG and z-cut quartz substrate, respectively, $d$ (= 1~nm) is the thickness of the BLG \cite{Gupta06a}, $\Delta L=-14~\mu$m is the thickness difference between sample and reference substrates (measured with a precision micrometer, and confirmed by THz-TDS data of the two bare substrates before BLG deposition), and $c$ is the speed of light in vacuum. This expression takes into account the multiple internal reflections inside the BLG sample, but does not include multiple reflections in the substrate --- we need not take substrate reflections into account because we have truncated the etalon pulse in our analysis. The complex refractive index ${\tilde n_{sub}}$ of z-cut quartz was first measured with vacuum as reference at different temperatures, obtained to be ${\tilde n_{sub}} \approx 2.11 + 0.002i$. This agrees with Ref.~\onlinecite{Loewenstein73}, showing that z-cut quartz is a very good THz transparent material with a temperature-independent, and almost frequency-independent, refractive index, in our frequency and temperature range. The complex refractive index $\tilde n(\omega ) = n(\omega ) + ik(\omega )$ is then extracted from Eq.~(\ref{eqn:Tw}) by numerical iteration, which is then used to calculate the complex optical conductivity $\tilde \sigma (\omega ) = {\sigma _1}(\omega ) + i{\sigma _2}(\omega )$, where ${\sigma _1}(\omega ) = 2nk\omega {\varepsilon _0}$ and ${\sigma _2}(\omega ) = ({\varepsilon _\infty } - {n^2} + {k^2})\omega {\varepsilon _0}$, ${\varepsilon _0}$ being the free space permittivity, and high frequency dielectric constant ${\varepsilon _\infty } = 8$ for graphene \cite{Kuzmenko08a}. However, the values of $\sigma_{2}(\omega)$ are very sensitive to the value of $\Delta L$, due to the very small thickness of BLG compared to $\Delta L$ ($\sim \mu$m). Hence we only discuss $\sigma_{1}(\omega)$ in our subsequent analysis. Note that, for a very thin metallic film on an insulating substrate, the following assumptions can be used: $n \gg n_{sub} > 1$ and $d \tilde{n} \omega/c \ll 1$, and Eq.~(\ref{eqn:Tw}) becomes the commonly-used thin-film expression \cite{Averitt02}
\begin{equation}
T(\omega) =
 \frac{1+ \tilde{n}_{sub}}{1+\tilde{n}_{sub}+Z_{0} \sigma (\omega) d} \exp [i \omega \Delta L({{\tilde n}_{sub}}-1)/c] \label{eqn:Twthinfilm}
 \end{equation} where $Z_{0}$ is the free space impedance. The values of $\sigma_{1}$ obtained from Eq.~(\ref{eqn:Twthinfilm}) are identical to that from Eq.~(\ref{eqn:Tw}).

\begin{figure} \centering \includegraphics[width=8cm,clip]{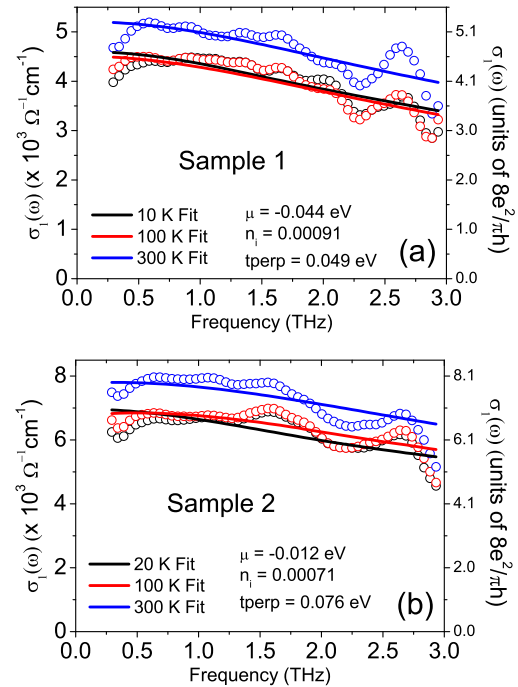}
\caption{(Color online) Real conductivity $\sigma_{1}(\omega)$ of BLG of (a) Sample 1 and (b) Sample 2. Circles = data. Solid lines = simultaneous fits of the unitary scattering model to the 10~K (black), 100~K (red) and 300~K (blue) data. Vertical axes on the right expresses $\sigma_{1}(\omega)$ in units of the minimum conductivity $8e^{2}/(\pi h)$ as specified in Nilsson's model \cite{Nilsson08a}.}
\label{fig:sigma}
\end{figure}

Figure~\ref{fig:sigma}(a) shows $\sigma_{1}(\omega)$ at 10~K, 100~K and 300~K of Sample 1. Note the very small difference between $\sigma_{1}(\omega,10~\text{K})$ and $\sigma_{1}(\omega,100~\text{K})$. Superposed on top of a Drude-like response, is a strong peak centered at $\sim$2.7~THz. We first analyze the Drude-like background using a theoretical model developed by Nilsson \textit{et al.} \cite{Nilsson08a} for unitary scatterers for Bernal BLG. The applicability of this theoretical model comes from the fact that  disorder broadens the low-energy features which would otherwise differentiate between the perfect Bernal-stacked BLGs and twisted BLGs.  Therefore, we anticipate the robust validity of this model for the analysis of the Drude-like background. The model starts by considering a Hamiltonian of the BLG under the tight binding model. Within the T-matrix approximation for unitary scattering, one derives the electron self-energy of BLG, which gives the Green's function in the presence of disorder via the Dyson equation. The conductivity is then calculated from the convolution of the Green's function elements (encoded in the kernal $\Xi$), as a function of chemical potential $\mu$, impurity concentration $n_{i}$, interlayer coupling (hopping) $t_{\perp}$ and temperature, to be
\begin{equation}
\sigma_{1}(\omega) \propto \frac{8e^{2}}{\pi h}\int d\epsilon \left [-\frac{n_{F}(\epsilon + \omega) - n_{F}(\epsilon)}{\omega} \right]\Xi (\epsilon, \epsilon + \omega),
\label{eqn:sigma}
\end{equation} where $n_{F}$ is the Fermi distribution function. The pre-factor $8e^{2}/(\pi h)$ is the \textit{approximate} minimal conductivity per BLG, whose exact value will depend on the actual distribution of impurities among the inequivalent sites of the $A$ and $B$ sublattices \cite{Nilsson08a}. In Fig.~\ref{fig:sigma}(a), the 10~K, 100~K and 300~K data were \textit{simultaneously} fitted with the model via Eq.~(\ref{eqn:sigma}), shown by black (10~K), red (100~K) and blue (300~K) solid lines. The resulting fitting parameters were $\mu^{fit} = -0.044$~eV, $n_{i}^{fit} = 0.00091$, and $t_{\perp}^{fit} = 0.049$~eV. Note the small difference between the 10~K and 100~K fits --- consistent with data. In fact, the theoretical $\sigma_{1}(\omega,100~\text{K})$ is smaller in magnitude than $\sigma_{1}(\omega,10~\text{K})$, showing that a spectral weight redistribution has taken place. We were initially surprised that the 100~K conductivity should be so similar to the 10~K conductivity, with the 300~K conductivity lying above both of them. These features, however, can be captured by the impurity-scattering model, lending strong credence to the validity of the model in explaining the THz data of Bernal BLG. In the above fittings, the fitted values of $\mu$ and $n_{i}$ are consistent with Raman data of the same sample on the same substrate \cite{EPAPS}, with charged impurity concentration $< 10^{13}$~cm$^{-2}$ (corresponding to $<0.0026$ per BLG) and $\mu = (-0.042 \pm 0.012)$~eV, determined by the Raman peak positions of the G and 2D bands, and the intensity ratio of the 2D and G peaks \cite{Das09,Ziegler11,Casiraghi09}. The unitary scatterer concentration of the sample is typically $<10^{-5}$, which was calculated using the ratio of the D and G peak intensities \cite{Cancado11}.

\begin{figure} \centering \includegraphics[width=8cm,clip]{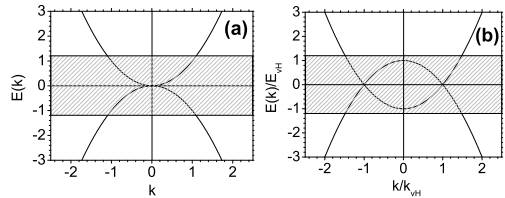}
\caption{Schematic dispersion of a (a) Bernal BLG and (b) twisted BLG. The shaded region indicates the states that have been broadened by disorder in the sample, hence our data for twisted BLG could be fitted with a theory developed for Bernal BLG, but with a smaller value of $t_{\perp}$.}
\label{fig:schematic}
\end{figure}

Figure~\ref{fig:sigma}(b) shows the frequency dependence of the real conductivity, $\sigma_{1}(\omega)$, at 20~K, 100~K and 300~K, of Sample 2. The \textit{simultaneous} fits (solid lines) of the data to Eq.~(\ref{eqn:sigma}) now yield fitting parameters $\mu^{fit} = -0.012$~eV, $n_{i}^{fit} = 0.00071$, and $t_{\perp}^{fit} = 0.076$~eV. Once again the fitted $\mu$ is consistent with Raman data, where $\mu = (-0.012 \pm 0.008)$~eV, and similar impurity concentration as Sample 1. Note that both fitted values of $t_{\perp}^{fit}$ are smaller than the value $t_{\perp}$=0.27~eV for Bernal BLG. In fact, for a single monodomain of twisted BLG, the interlayer hopping is angle dependent, but for small angles it can be approximated as  $t_{\perp}^{\theta} \approx 0.4 t_{\perp} \approx 0.1$~eV \cite{Santos12}, and $t_{\perp}^{\theta}<$0.1~eV for larger $\theta$'s (larger $\theta$ implies a larger separation between the layers, hence smaller interlayer hopping). Hence the value of $t_{\perp}^{fit}$ obtained ($\approx$50--70~meV) could be an average of interlayer couplings from all possible twisting angles in the sample, re-expressed in the form of perfect Bernal stacking. The fit to a theory based on an "effective" Bernal BLG only works because disorder broadens all the features which would otherwise distinguish Bernal from twisted BLG at low energies, namely, the presence of VHS in the density of states of twisted BLGs \cite{Santos07a} (see Fig.~\ref{fig:schematic}). The difference in absolute values of $\sigma_{1}(\omega)$ between samples 1 and 2 is consistent with sample-to-sample variations observed in other works on BLG \cite{DasSarma2010}.

A close inspection of $\sigma_{1}(\omega)$ reveals the presence of a peak that appear on top of the background signal. In twisted BLG, VHS develop near the Fermi energy, which results in an enhanced density of states \cite{Santos07a}. The energy scale of such VHS depends sensitively on the twisting angle $\theta$, given by
\begin{equation}
E_{vhs} = \frac{8 \pi \hbar v_F}{3a} \left|\sin\left(\theta/2 \right)\right| - 2 t_{\perp}^{\theta},
\label{eqn:Evhs}
\end{equation} where $v_{F} = 1.0 \times 10^6$ m/s \cite{Heer07a} is the Fermi velocity and $a$$\approx$2.46~\AA~the lattice constant. We observed a strong peak at $\sim$2.64~THz, whose presence is reproducible from sample to sample. Its position is consistent with the second non-zero $E_{vhs}$, computed from Eq.~(\ref{eqn:Evhs}) to be 2.77~THz, and corresponds to $\theta_{28} = 1.161^{\circ}$ (from Eq.~(\ref{eqn:theta})). Note that the first non-zero $E_{vhs}$ of 0.89~THz, arising from $\theta_{29} = 1.121^{\circ}$, is not visible from Fig.~\ref{fig:sigma}. Theoretical density of states calculations \cite{Santos12} show that, for $\theta_{27}$=1.20$^{\circ}$, the van Hove peaks are still barely visible, whereas for $\theta_{30}$=1.08$^{\circ}$ the VHS have disappeared \cite{Santos12}. Also, since disorder builds up the density of states near the Dirac point, the VHS are broadened by being in the middle of a continuum of disordered states. These factors may explain our inability to see any clear feature near 1~THz. Note that our 2.7-THz peak is robust against the type of windowing function we used before performing FFT. Besides the conventional windowing functions, we also constructed an asymmetric windowing function that is tailored to the shape of our asymmetric time-domain waveforms \cite{Galvao07} --- all yielded the 2.7-THz peak. This 2.7-THz ($\sim$11~meV) $E_{vhs}$ is also consistent with scanning tunneling microscopy and spectroscopy (STM/STS) works on twisted BLG \cite{Li10a,Yan12}.

The Raman data on the same samples also gave information about the twisting \cite{EPAPS}. The position of the $G$ peak shows the samples to be slightly $p$-doped \cite{Casiraghi09}. The blueshift of the $2D$ peak implies the existence of twisting, and the value of the $2D$ peak width implies a twisting angle $\theta < 5^{\circ}$ \cite{Kim12}. The consistency of these results across different positions of the samples implies a well defined twisting angle in our samples. Hence our THz data, besides being consistent with Raman data in the same samples, points out the exact twisting angle, and shows the effect on twisting on the optical conductivity.

In conclusion, we have studied the far-infrared dielectric response of bilayer graphene at different temperatures by THz-TDS. On top of a Drude-like response, we observed a peak in the real part of optical conductivity. The overall Drude shape was analyzed using a disorder-dependent model, while the conductivity peak at 2.7~THz was attributed to an enhanced density of states at that energy, that is caused by the presence of a low-energy van Hove singularity arising from a commensurate twisting of the top graphene layer relative to the bottom layer. A unified theory that considers the effect of both disorder and twisting on BLG conductivity is clearly desired.

We thank J. Nilsson, J. M. B. Lopes dos Santos, N. M. R. Peres, E. Y. Andrei and R. D. Averitt for useful discussions. E.E.M.C. acknowledges support from Singapore MOE AcRF Tier 2 (ARC 23/08), as well as the NRF CRP (NRF-CRP4-2008-04). J.-X.Z. is supported by the NNSA of the U.S. DOE at LANL under Contract No. DE-AC52-06NA25396 and the U.S. DOE Office of Basic Energy Sciences. A. H. C. N. acknowledges NRF-CRP award ``Novel 2D materials with tailored properties: beyond graphene'' (R-144-000-295-281), DOE grant DE-FG02-08ER46512, and ONR grant MURI N00014-09-1-1063.

\bibliography{Graphene1}

\end{document}